\documentstyle[prl,aps,amssymb,epsfig]{revtex}

\begin{document}
\draft

\twocolumn[\hsize\textwidth\columnwidth\hsize\csname@twocolumnfalse\endcsname

\title{Manifestation of multiband optical properties of MgB$_2$}

\author{A.~B.~Kuz'menko$^{1}$, F.~P.~Mena$^{1}$, H.~J.~A.~Molegraaf$^{1}$,
D.~van~der~Marel$^{1}$, B.~Gorshunov$^{2}$, M.~Dressel$^{2}$,
I.~I.~Mazin$^{3}$, J.~Kortus$^{4}$, O.~V.~Dolgov$^{4}$,
T.~Muranaka$^{5}$, and J.~Akimitsu$^{5}$} \address{$^{1}$Material
Science Center, University
 of Groningen, Nijenborgh 4, 9747 AG Groningen, The Netherlands}

\address{$^{2}$1. Physikalisches Institut, Universit\"at Stuttgart,
Pfaffenwaldring 57, D-70550 Stuttgart, Germany}

\address{$^{3}$Center for Computational Materials Science, Code 6390,
Naval Research Laboratory, Washington, DC 20375}

\address{$^{4}$ Max-Planck-Institut f{\"u}r Festk{\"o}rperforschung,
D-70569, Stuttgart, Germany}

\address{$^{5}$Department of Physics, Aoyama-Gakuin University, 6-16-1
Chitsedai, Setagaya-ku, Tokyo 157, Japan} \date{\today} \maketitle

\begin{abstract}

The optical conductivity of MgB$_2$ has been determined on a dense
polycrystalline sample in the spectral range 6 meV - 4.6 eV using
a combination of ellipsometric and normal incidence reflectivity
measurements. $\sigma_{1}(\omega)$ features a narrow Drude peak
with anomalously small plasma frequency (1.4 eV) and a very broad
"dome" structure, which comprises the bulk of the low-energy
spectral weight. This fact can be reconciled with the results of
band structure calculations by assuming that charge carriers from
the 2D $\sigma$-bands and the 3D $\pi$-bands have principally
different impurity scattering rates and negligible interband
scattering. This also explains a surprisingly small correlation
between the defect concentration and $T_c$, expected for a two-gap
superconductor. The large 3D carrier scattering rate suggests
their proximity to the localization limit.

\end{abstract}
]

\vskip2pc

%

The recently discovered copperless superconductor MgB$_{2}$\cite
{Nagamatsu} has attracted substantial attention. Most researchers
agree that MgB$_{2}$ is a conventional $sp$-metal, where a
combination of strong bonding and sizable density of states
produces a high critical temperature within the standard theory
\cite{Kortus,Pickett,Kong}. There is an agreement among theorists
that there are two qualitatively different systems of bands in
MgB$_{2}$: quasi-2D $\sigma $-bands, and 3D $\pi $-bands, and that
the former strongly couple to optical E$_{2g}$ phonon at $\approx$
600 cm$^{-1}$. There is also experimental \cite{Buzea} and
theoretical\cite{Liu} evidence of two different superconducting
gaps in this compound.

Optical spectroscopy is a direct probe of the electronic
structure, and it has not been supportive of the conventional
picture so far: the reported plasma frequency \cite{Kaindl,Tu} is
$\approx$ 1.5 eV, in utter disagreement with LDA calculations of
$\approx$ 7 eV \cite{Kortus,Kong,Liu}. It is tempting to ascribe
this to correlation effects, so that the missing spectral weight
is shifted to some incoherent excitations. However, there is no
reason to expect strong correlations in MgB$_2$, which does not
have any d- or f- elements, and is in no proximity to magnetism.
Furthermore, recent angle-resolved photoemission data on a single
crystal \cite{Uchiyama} have found a remarkable correspondence of
the observed bands to LDA calculations, implying that there is
virtually no renormalization of Fermi velocity in this system, as
opposed, for instance, to high-T$_c$ cuprates. Different values
for the carrier scattering rate at $T\sim T_{c}$ have been
reported from transmission spectroscopy on thin films: 150
cm$^{-1}$ (18.6 meV)\cite{Pronin}, 300 cm$^{-1}$ (37.2 meV)
\cite{Kaindl} and 700 cm$^{-1}$ (86.8 meV) \cite {JungOptics}.
Lacking large single crystals, a study of polycrystalline samples
is justified as a first step to understand optical properties of
this material. In this Communication we report the optical
conductivity of MgB$_{2}$ obtained from a dense polycrystalline
sample from ellipsometry and normal-incidence reflectivity in a
range from 6 meV to 4.5 eV, which comprises both intraband and
low-energy interband excitations. We also report theoretical
calculation of the same quantity, including the electron-phonon
interaction (EPI) and the polycrystalline nature of the sample. We
suggest that the intraband conductivity consists of two
qualitatively different components: a narrow Drude peak (DP) due
to the $\sigma$-bands, and an overdamped DP due to the $\pi$-band
electrons. The latter scatter so strongly in our sample that one
may expect deviations from the Drude shape due to localization
effects, and indeed we find that the feature in question deviates
from the Drude formula. This model explains naturally a yet
unresolved paradox of the two-band scenario \cite{Liu,Shulga}: the
absence of a pair-breaking by paramagnetic impurities, predicted
for two-gap superconductors \cite{Golubov97}. In this
Communication we concentrate on the broad-range optical properties
keeping the FIR study of the superconducting gap for a separate
publication.

\begin{figure}
    \centerline{\psfig{figure=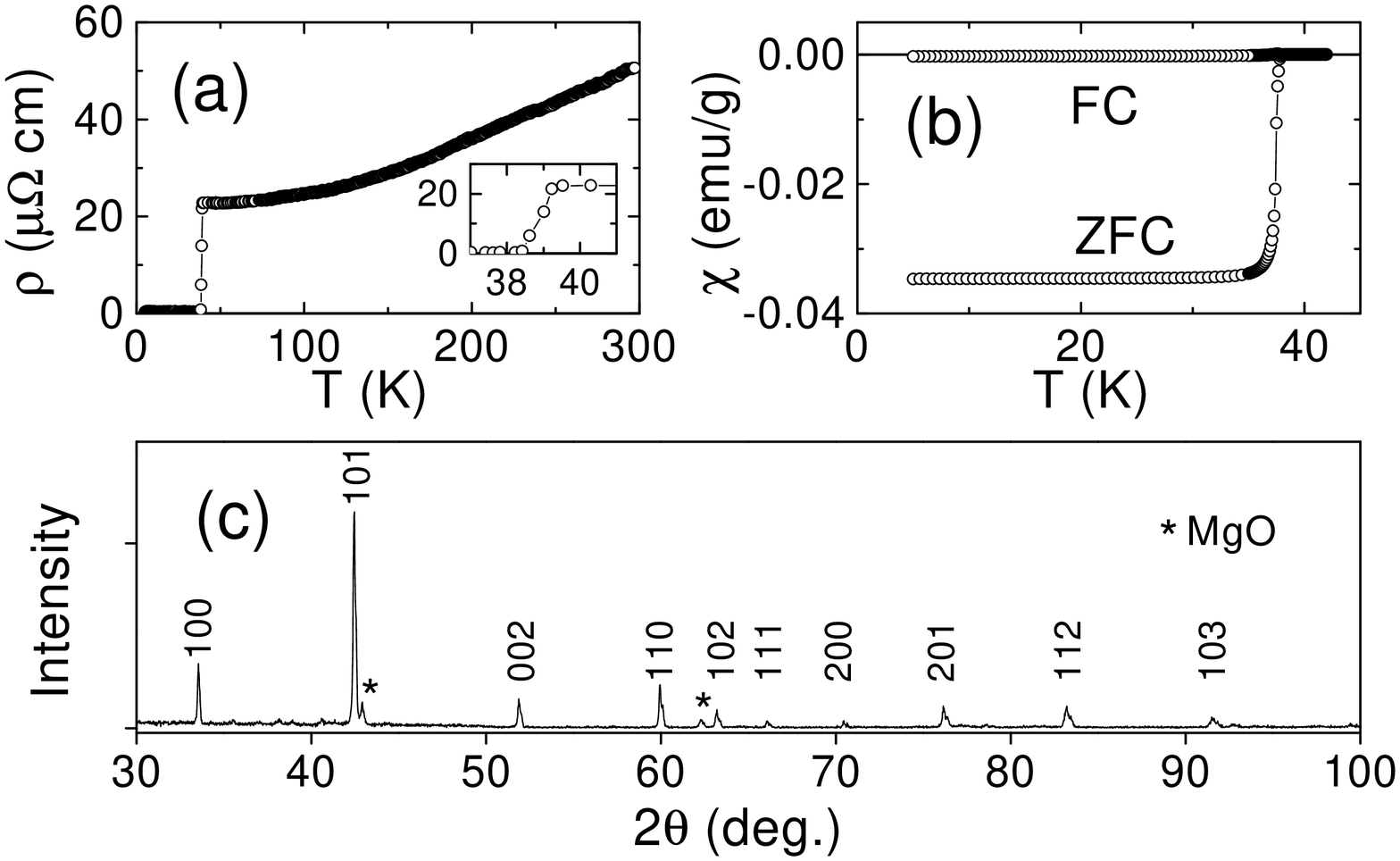,width=8.5cm,clip=}}
    \caption{The DC resistivity (a), the magnetic susceptibility (b) and the X-ray
     diffraction (c) of the MgB$_{2}$ sample.}
     \label{Res}
\end{figure}

The sample was synthesized from high-purity powdered Mg and B, as
described elsewhere \cite{Nagamatsu}, and then treated in a
high-pressure BN cell at 1623 K and 5.5 GPa for 30 minutes and
finally quenched to room temperature. A very dense metallic-like
sample was obtained. The $T_{c}$ is about 39 K with transition
width of less than 1 K both from the DC resistivity and the
magnetic susceptibility (Fig.\ref{Res} (a, b)). The estimated
superconducting volume fraction is about 60 \%. The X-ray
diffractometry (Fig.\ref{Res}(c)) revealed presence of impurity
phases which is difficult to avoid in a bulk polycrystalline
sample. One can estimate that the sample contains some 5 - 7 \% of
MgO and about 1 \% of other phases. The R(300K)/R(50K) ratio is
2.2.

The sample was cut and dry-polished with a 0.1 $\mu $m diamond
micropolisher. The surface was black and shiny with an area of
about 8 mm$ ^{2}$. Because of ellipsometrically detected changes
of the optical properties in air, probably due to surface
oxidation, the sample was inserted into the cryostat within
minutes after polishing to make sure that the near-surface region
corresponds to the bulk properties. Normal-incidence reflectivity
measurements in polarized light did not reveal any optical
anisotropy implying no preferential grain orientation.

We combined ellipsometry and reflection spectroscopy to determine
the optical constants for 15 K$<$ $T$ $<$ 300 K. The dielectric
function $\epsilon =\epsilon _{1}+i\epsilon _{2}$ was measured
directly using a commercial optical ellipsometer with a UHV
cryostat in the range 6000 - 37000 cm$^{-1}$ (0.45 - 4.6 eV) at
the angle of incidence $\theta$ = 80$^\circ$. The measurements at
$\theta$ = 60$^\circ$ gave almost the same dielectric function,
which is also an indication of isotropic properties of our sample.
The normal incidence reflectivity spectra were measured from 20 to
6000 cm$^{-1}$ (2.5 meV - 0.45 eV) utilizing Bruker 113v FT-IR
spectrometer, equipped with a home-design ultra-stable optical
cryostat. Absolute reflectivities for each temperature were
calibrated with the aid of a gold film, deposited {\it in situ} on
the sample. In this range the complex reflectivity phase was
obtained with Kramers-Kronig (KK) transformation, allowing us to
calculate $\epsilon _{1}$ and $\epsilon _{2}$. The phase spectrum
was anchored using ellipsometric data at higher frequencies (see
Ref.\cite{vdEb}).

\begin{figure}
    \centerline{\psfig{figure=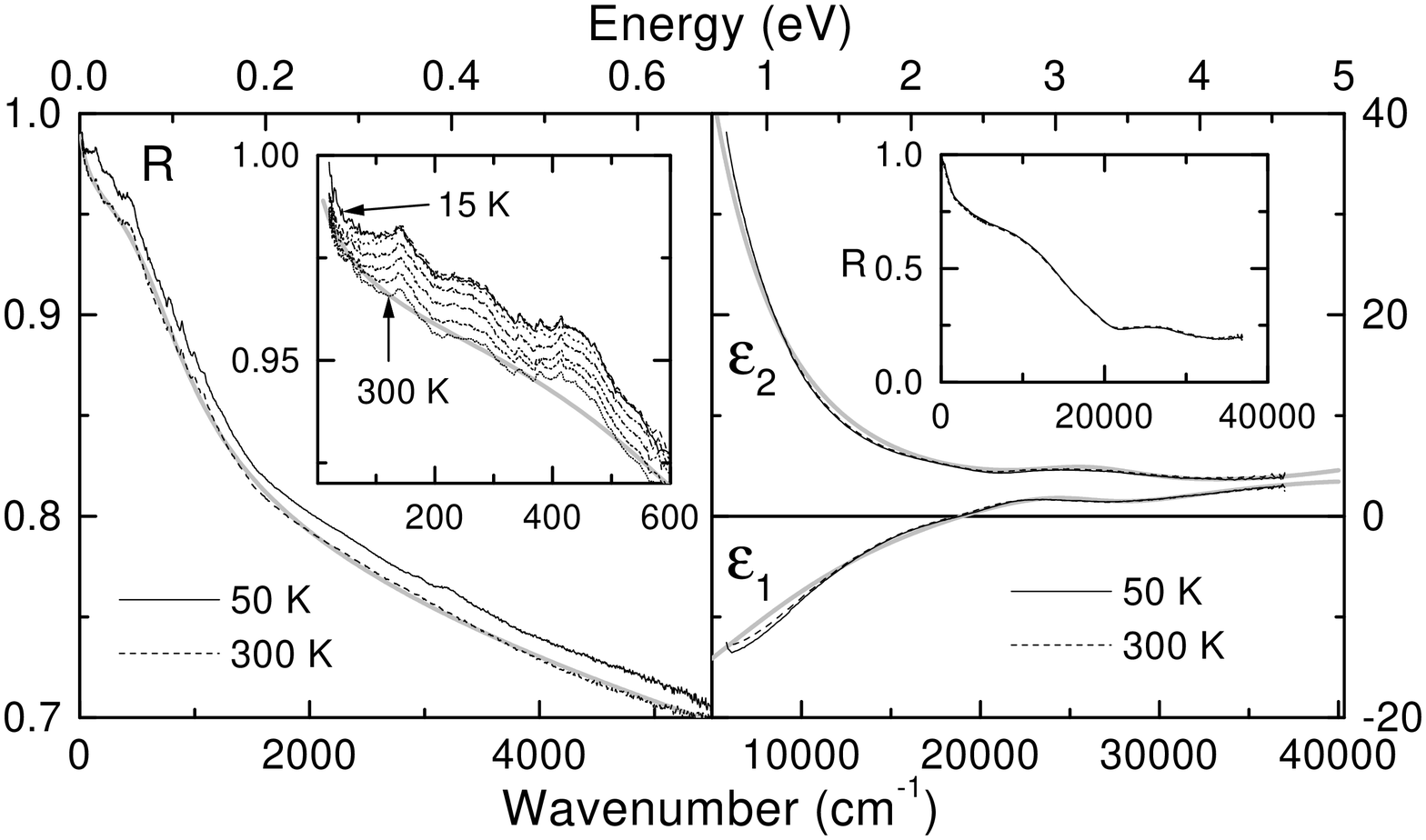,width=8.5cm,clip=}}
    \caption{Experimental spectra at 50 K and 300 K: normal reflectivity
    (left) and dielectric function (right). The left inset shows the
    reflectivity in the FIR range for $T=$300, 250, 200,
    150, 100, 50 and 15 K (from bottom to top). The right
    inset shows the reflectivity in the full frequency range (at high frequencies,
    restored from ellipsometry data). The Drude-Lorentz
    fit at 300 K is shown in gray.}
    \label{exp}
\end{figure}

An important issue is the effect of impurity phases on the optical
spectra. The main impurity component is MgO, which has almost zero
optical conductivity in the studied frequency range except for a
few sharp phonon peaks in the FIR range. Using the Maxwell-Garnett
theory of the two-component media \cite{Stroud75} one can show
that the role of impurity in this case is to reduce the observed
conductivity uniformly by 7 - 10 \% compared to the impurity-free
sample. As impurity concentration is not precisely known, we do
not correct the conductivity for this effect, but keep it in mind
in the discussion.

\begin{figure}
    \centerline{\psfig{figure=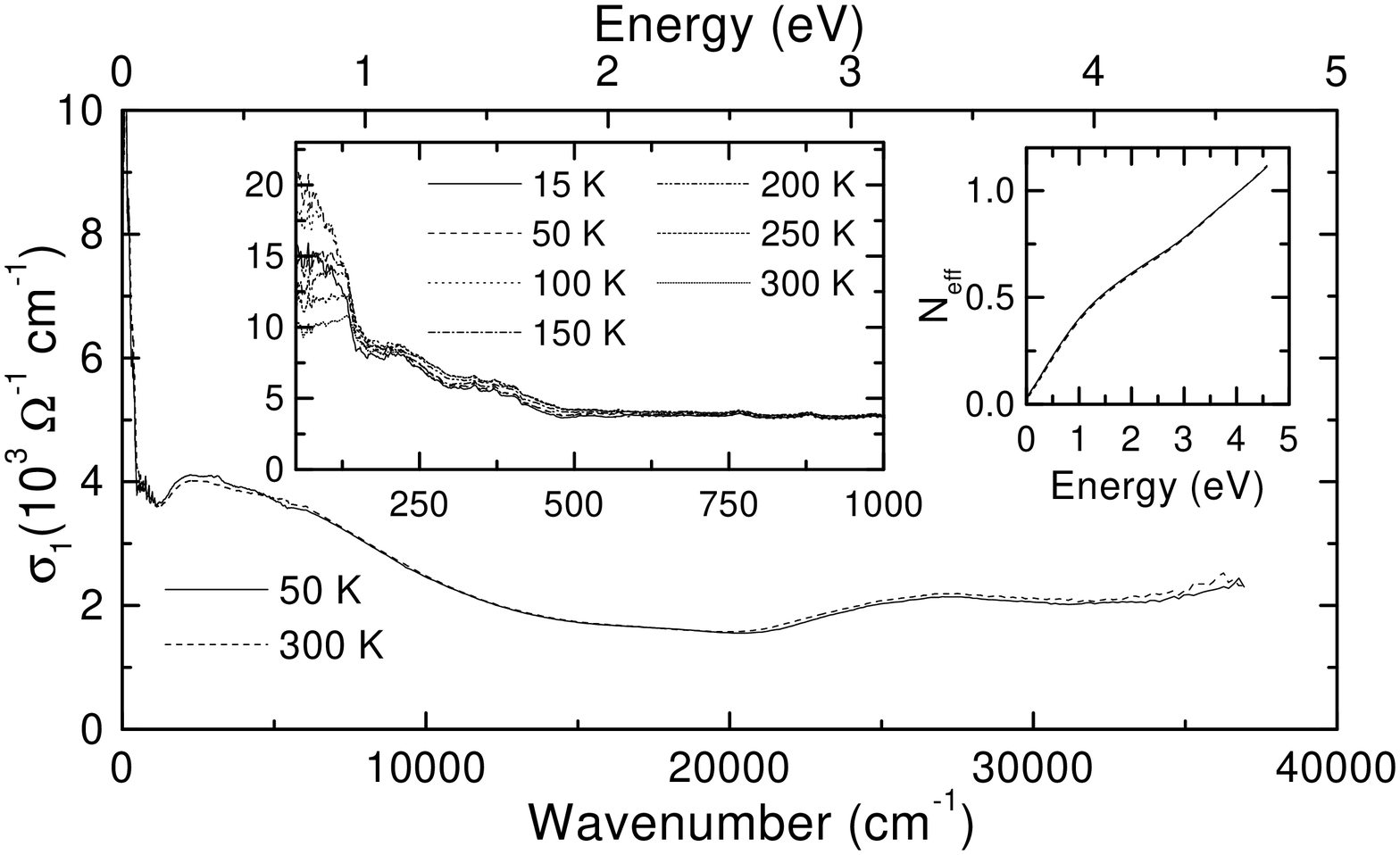,width=8.5cm,clip=}}
    \caption{Optical conductivity at 50 K and 300 K determined from the
    experimental data (as described in the text). The left inset shows
the $T$-dependence of the narrow Drude peak is shown. The
    right inset: the effective number of carriers per cell, $N_{\mbox{\scriptsize eff}}(\omega)$ (solid
curve:  50 K,
    dashed curve: 300 K). }
     \label{cond}
\end{figure}

The experimental spectra are shown in Fig.\ref{exp}. At high
energies the spectra demonstrate very little $T-$%
dependence. In the IR range the reflectivity steadily increases
with cooling down. An additional small superconductivity-related
increase of $R$ is observed below $T_{c}$ for $\omega $ $\lesssim$
12 meV, in agreement with earlier data \cite{Gorshunov}. The main
low-frequency features of the optical conductivity
(Fig.\ref{cond}) are a narrow DP with a small spectral weight, and
a broad "dome" structure extending up to 1 - 1.5 eV. There is a
minimum at $\approx$ 0.15 eV and a broad maximum at $\approx$ 0.3
eV. The structure itself looks like a strongly damped second DP.
Another broad peak is centered at about 3.3 - 3.4 eV. One can note
some phonon-like structures in the FIR reflectivity and,
correspondingly, conductivity curves. Those structures are not
reproducible from sample to sample, therefore we relate most of
them to impurity phases \cite{ftnt}.

The standard procedure of quantifying optical spectra is the
Drude-Lorentz data fitting: $\epsilon({\omega
})=\epsilon_{\infty}-\Omega _{p}^{2}/(\omega (\omega +i\gamma ))
+\sum_{i}\Omega _{pi}^{2}/(\omega _{0i}^{2}-\omega ^{2}-i\gamma
_{i}\omega )$, where $\Omega _{p}$ and $\gamma $ are the plasma
frequency and the scattering rate for the DP, and $\omega _{0i}$,
$\Omega _{pi}$ and $\gamma _{i}$ are the frequency, the "plasma"
frequency, and the scattering rate of the $i$-th Lorentz
oscillator (LO), $\epsilon_{\infty}$ is the high-frequency
dielectric constant. We fitted simultaneously $R(\omega )$ at low
frequencies and both $\epsilon _{1}(\omega )$ and $\epsilon
_{2}(\omega )$ at high frequencies (see Fig. \ref{exp}) with one
set of parameters listed in the Table for 300 K.

\begin{figure} \centerline{\psfig{figure=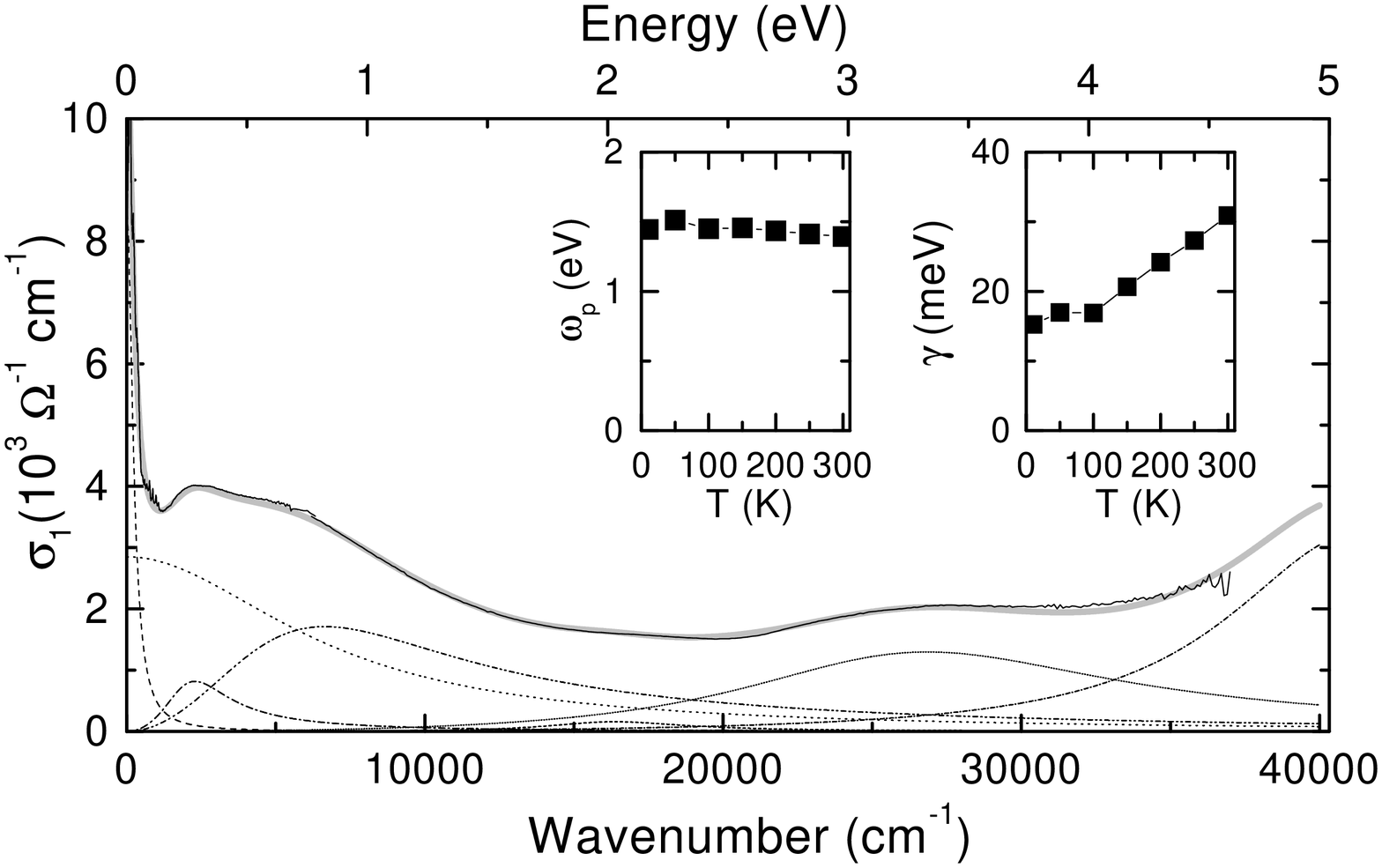,width=8.5cm,clip=}}
\caption{Contributions to the optical conductivity in the
Drude-Lorentz fitting at 300 K. The solid line is the experiment,
the thick gray line is the fit; the dashed (dotted) lines show the
narrow (broad) Drude peak and all the other curves show different
LOs. The insets: temperature dependence of the $\Omega_p $
and $\gamma$ for the narrow Drude peak.}
\label{sosc} \end{figure}

A good fit can be obtained by introducing two DPs
with very different properties (Fig.\ref {sosc}). The narrow DP
has an almost temperature-independent plasma frequency $\approx$
1.4 eV; the temperature dependence of the scattering rate (see
inset in Fig.\ref{sosc}) scales well with the
$\rho_{\mbox{\scriptsize DC}}(T)$ curve (Fig.\ref{Res}(a)). The
second DP is anomalously broad ($\gamma$ $\approx$ 1.1 - 1.2 eV).
Although its plasma frequency ($\approx$ 4.9 eV) is much larger
than that of the narrow peak, the latter determines the major part
of the static conductivity. The "dome" structure
can be fitted with a combination of a broad DP and two LO at
$\sim$ 0.3 and $\sim$ 0.8 eV. Finally, there is another broad peak
centered at $\sim$ 3.4 eV and a peak at $\sim$ 5.1 eV.

Let us now compare the data with the first-principles calculations
\cite{Liu,Kortus}. To account for the polycrystalline sample
texture we averaged the in-plane and out-of-plane contributions to
the optical conductivity using the effective medium approximation
(EMA) \cite{Stroud75} assuming spherical randomly oriented grains.
This approach is more accurate at low frequencies, where the
wavelength $\lambda$ is larger than the typical grain size $d$ (in
our case of the order of 1-5 $\mu$m). For the limit $\lambda \gg
d$ the reflectivity should be averaged directly rather than
conductivity. For highly anisotropic compounds, e.g. cuprates, it
was shown that some artificial structures may show up due to
reflectivity-averaging effect \cite{Orenstein88}. Although we take
such a possibility very seriously, the fact that the complex
dielectric functions obtained by ellipsometry measurements at
incidence angles 60$^\circ$ and 80$^\circ$ match very well is an
argument that the concept of effective medium conductivity works
sufficiently well even in the high-energy part of our measured
spectral range.

\begin{figure}
    \centerline{\psfig{figure=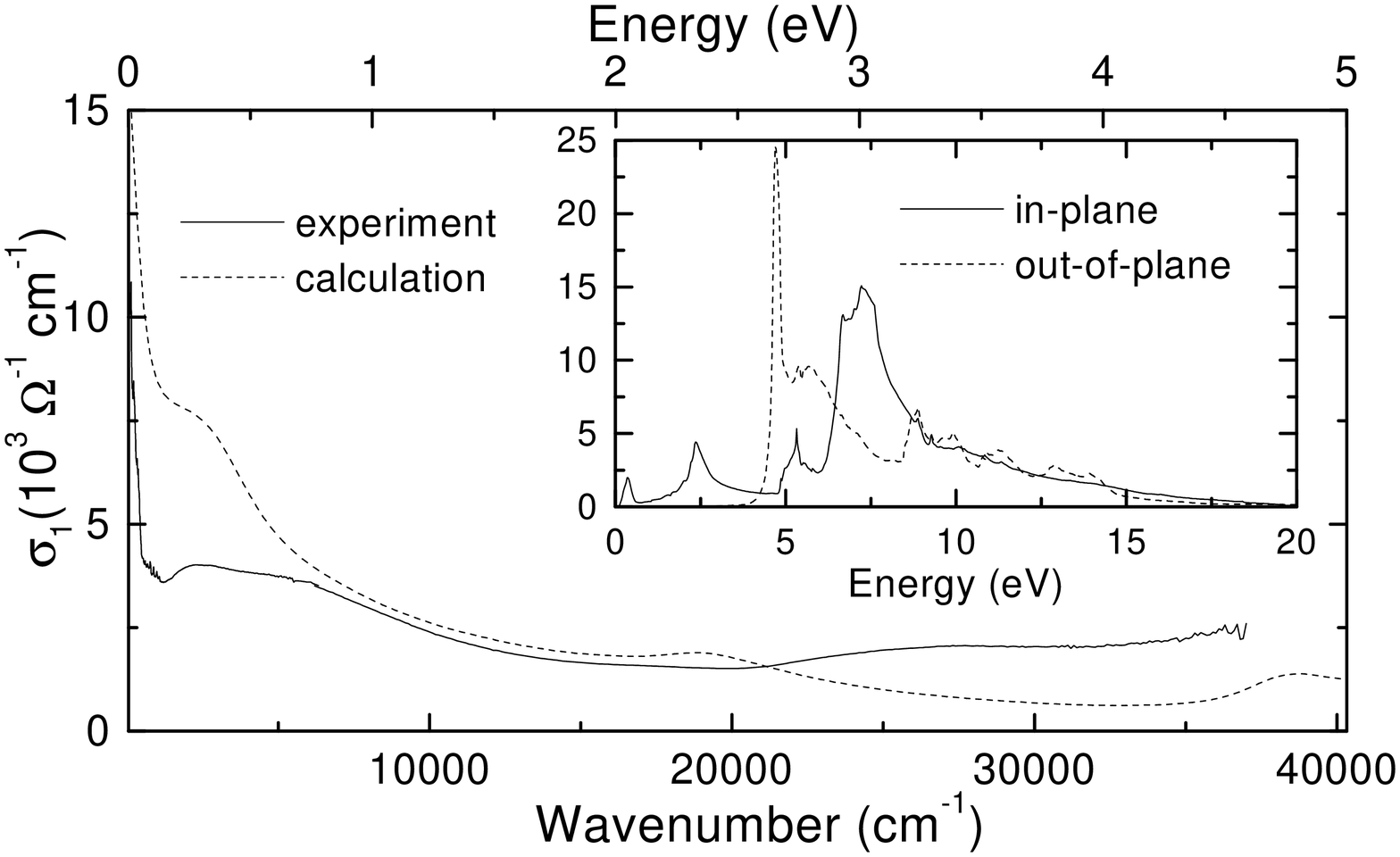,width=8.5cm,clip=}}
    \caption{Comparison of the experimental and calculated
 optical conductivity, including the interband part shown
in the inset, at 300 K (see text).
}
     \label{theory}
\end{figure}

The interband optical conductivity (Fig.\ref{theory}) was
calculated by the WIEN97 LAPW code \cite{Optic} using the band
structure of Ref.\cite{Kortus}. The experimental peak at 0.3 eV
clearly corresponds to the in-plane interband transition between
the two $\sigma$-bands at 0.35 eV. The broad 3.4 eV peak finds no
counterpart in the theory: the closest peak (2.4 eV, in-plane
polarization) stems from the van Hove singularity at the M point.
The reason for this discrepancy is not clear at the moment,
although it might be related to the disorder effects discussed
below. The increase of the conductivity above 4 eV, which is
simulated by the LO at 5.1 eV, one can probably associate with the
intensive out-of-plane maximum at 4.7 eV due to charge excitations
between the B and Mg planes \cite{Ku}.

The intraband conductivity is particularly interesting. The calculated
in-plane (out-of-plane)
plasma frequencies of the 2D $\sigma$-band are 4.1 (0.7) eV, which
in the EMA gives 3.0 eV for the effective plasma frequency
$\omega_{p\sigma}$ of the $\sigma$-band. Similarly, for the
$\pi$-band, we obtain $\omega_{p\pi}$ = 6.2 eV. The total plasma
frequency $\omega_{p}$ is thus expected to be about 6.9 eV.

The observed $\Omega_{p}$ of the narrow Drude peak alone (1.4 eV),
in agreement with Refs.
\cite{Kaindl,Tu}, is much smaller than the calculated total
$\omega_{p}$. It can account for only a small part of the expected
total spectral weight. However, about 80\% of the missing spectral
weight can be found in the features forming the broad "dome"
structure. We suggest therefore that the narrow DP is due to the
2D $\sigma$-band carriers, while the broad "dome" is mainly formed
by the intraband excitations within the 3D $\pi$-band with some
contribution from the low-energy interband transitions between
$\sigma$-bands. The two orders of magnitude difference in the
elastic scattering rates for the $\sigma$- and $\pi$-bands can be
explained by the principally different nature of the two band
systems. The $\sigma$-bands have hardly any weight on Mg, while
the $\pi$-bands hybridize with Mg orbitals. On the other hand, the
strong covalent bonding in the B planes makes any defect in this
plane energetically very unfavorable, while the loosely bound Mg
planes should be prone to all kinds of defects. This implies that
electrons in the $\sigma$-bands may experience much less defect
scattering than those in the $\pi$-bands.

Even this model cannot fully explain the small plasma
frequency (1.4 eV) of the narrow DP, since the calculated
$\sigma$-band effective plasma frequency is 3 eV. However, an
important piece of physics is the mass renormalization due to
strong EPI. This effect is negligible if $\gamma \gg \omega
_{ph}$, which is usually the case in metal optics. In MgB$_{2}$,
however, the EPI is believed to come mainly from the E$_{2g}$
phonon with $\omega _{ph}\sim 600$ cm$^{-1},$ larger than the
$\sigma$-band scattering rate. In this case, the mass
renormalization still does not change the DC resistivity, but the
peak gets narrower, and fitting by the standard Drude formula
produces an $\omega _{p}$, which is, compared to the bare plasma
frequency, reduced by a factor $m^{*}/m = 1 + \lambda$, where
$\lambda$ is the EPI constant \cite{Dolgov95}. The remaining
spectral weight is transferred to higher energies, $\omega >
\omega _{ph}$. The calculated
coupling constants\cite{Liu} for different bands are: $\lambda _{0\sigma }$
= 1.2 and $\lambda _{0\pi }$ = 0.45. This
renormalizes the plasma frequency of the $\sigma$-band DP to $\approx 2$
eV.

Given such a large scattering rate, is the Drude model applicable
at all to the $\pi$-bands? The calculated Fermi velocity is about
4.9 $\times $ 10$^{7}$ cm/s \cite{Kortus} so that the $\gamma $
=1.1 eV =2.7 $\times$ 10$^{14}$ s$^{-1}$, results in a mean free
path $l$ = $v_{F}/\gamma$ $\approx$ 18 \AA , comparable with
 the lattice constant. This signals approaching weak
localization of charge carriers by defects, as opposed to
Boltzmann transport. The grain boundaries might introduce
additional scattering, however, such a short mean free path is
difficult to explain by the grain effects only. Localization
effects should lead to deviations from the pure Drude formula: an
additional suppression of conductivity at $\omega \rightarrow 0$
and a characteristic maximum on $\sigma_{1}(\omega)$ curve at
frequencies of the order of the inelastic scattering rate. This
might explain the broad peak at 0.8 eV, which has almost the same
$\gamma$ as the broad DP, and comparable spectral weight (see the
Table). The combination of these two terms might mimic the
deformation of the conventional DP due to localization effects.

Fig.\ref{theory} shows the effective conductivity (including both
interband and intraband components), obtained from first-principle
calculations. We calculated the intraband absorption separately
for the $\sigma-$ and $\pi-$bands, and included the
frequency-dependent mass renormalizations and scattering rates due
to EPI, as described in Ref.\cite{Dolgov95}. The EPI Eliashberg
function was taken in the Einstein form: $\alpha
_{\sigma,\pi}^{2}F(\omega )=0.5\lambda _{0\sigma,\pi}\omega \delta
(\omega -\omega _{ph})$. The values for the $T$ = 0 scattering
rates were taken from the experiment: $\gamma_{\sigma}$= 15 meV,
$\gamma_{\pi}$ = 1.1 eV.

There is a qualitative agreement between the first-principles
calculations and the experiment, despite a 50 \% difference in
plasma frequencies. However, without the two-band scenario the
discrepancy would be an order of magnitude bigger. The remaining
error may be due to one or more of the following: above mentioned
7 - 10 \% scaling of the conductivity due to finite impurity
concentration; surface quality issues; grain boundary and
localization effects; possible underestimation of $\lambda$ in the
calculations because of nonlinear EPI.

Let us now make a link to superconductivity. Various experiments
indicate that MgB$_2$ cannot be described by the standard BCS
theory\cite{Buzea}. The popular two-gap theory \cite{Shulga,Liu}
seems to be able to explain the majority of the experiments.
However, it is known \cite{Golubov97} that in this case the
interband impurity scattering must rapidly suppress $T_{c}$ by
isotropization of the order parameter. Such a suppression has not
been observed\cite{Buzea}, which can be interpreted as a serious
argument against the two-gap model. Our two-Drude picture of the
low-energy optical conductivity helps to resolve this paradox.
Indeed, the existence of two distinguishable Drude peaks would
only be possible provided that the $\sigma -\pi$ charge scattering
is small. In this case impurity scattering, albeit reducing the DC
conductivity, does not isotropize the gap, therefore the Anderson
theorem holds and no pairbreaking occurs \cite{Golubov97}.

It is worth noting that one cannot, based just on the fit quality,
choose between our ``two-DP'' model and a DL fit with the second
DP replaced by a broad low-energy ($\omega_{0} \alt $ 0.25 eV) LO.
Although the latter interpretation is consistent with the data, we
favor the  two DP model for the following reasons: (i) the plasma
frequency, unlike some other electronic properties, is known to be
reproducible by LDA calculations within errors hardly ever larger
than 100 \%, as opposed to a factor of five discrepancy we would
have for MgB$_{2}$. (ii) With one DP the total $\omega_p$ is less
than 2 eV; in this case the application of the standard high-$T$
limit of the Bloch-Gr\"uneisen formula, $\rho_{\mbox{\scriptsize
DC}}(T)=8 \pi^2 \lambda_{tr} T/ \omega_p^2$ would give the EPI
transport constant $\lambda_{tr} < 0.15$ \cite{Tu}, an
unphysically small value for an covalent-ionic metal with a
sizeable density of states. Finally, (iii) the two-DP model
resolves naturally the main problem of the two-gap
superconductivity model, supported by a variety of experiments.

In summary, the optical conductivity of MgB$_{2}$ in a wide
frequency range was obtained using a combination of ellipsometry
and normal incidence reflectometry methods. The drastic
suppression of the low-energy spectral weight finds a natural
explanation in the multiband model: a narrow Drude peak
corresponds to the 2D $\sigma$-bands (additionally renormalized by
the EPI), while the intraband absorption of the 3D $\pi$-bands is
spread over a large frequency range up to 1 - 1.5 eV. Optical
measurements on high-quality single crystals of MgB$_{2}$ are
necessary to get more reliable data on the optical conductivity
and its anisotropy.

This investigation was supported by the Netherlands Foundation for
Fundamental Research on Matter (FOM) with financial aid from the
Nederlandse Organisatie voor Wetenschappelijk Onderzoek (NWO). We
thank R. K. Kremer for performing the succeptibility measurements.

\begin{table}[tbp]
\caption{Parameters (in eV) of Drude (1, 2) and Lorentz (3 - 6)
peaks obtained by spectra fitting at 300 K ($\epsilon_{\infty}$ =
2.9).}
\begin{tabular}{cccccccc}
No.&$\omega_{0}$ & $\Omega_{p}$ & $\gamma$ & No.& $\omega_{0}$ &
$\Omega_{p}$ & $\gamma$ \\ \cline{1-4} \cline{5-8}\\ 1   &   - &
1.39    &   0.03    &   4   & 0.79 & 3.35 &   1.21    \\ 2   & - &
4.94    &   1.16 & 5 & 3.40 & 4.85    &   2.40    \\ 3   & 0.29 &
1.46    & 0.34 & 6 & 5.23 &   5.80    &   1.23    \\
\end{tabular}
\label{TabParams}
\end{table}


\begin{references}
\bibitem{ftnt}
The measured $\sigma_{1}(\omega)$ extrapolated to zero frequency
is about twice smaller than that from direct DC measurements. It
can be explained by the well-known uncertainty of the KK method at
very low frequencies, where reflectivity approaches unity. For
instance, shifting our spectra by 1\% up, which reflects a typical
systematic error of this kind of measurement, would make the
static limit agree well to the DC conductivity, with only minor
changes at high frequencies.
\end{references}
\end{document}